\documentclass[letterpaper, 10 pt, conference]{ieeeconf}  

\pdfoutput=1
\IEEEoverridecommandlockouts

\usepackage{graphics} 
\usepackage{epsfig} 
\usepackage{mathptmx} 
\usepackage{times} 
\usepackage{amsmath} 
\usepackage{amssymb}  
\usepackage{xcolor}

\usepackage{tikz}
\usepackage{lipsum}

\newcommand\copyrighttext{%
  \footnotesize 978-1-7281-1179-7/21/\$31.00~\copyright2021 IEEE | DOI: 10.1109/EMBC46164.2021.9630217.  Personal use of this material is permitted. Permission from IEEE must be obtained for all other uses, in any current or future media, including reprinting/republishing this material for advertising or promotional purposes, creating new collective works, for resale or redistribution to servers or lists, or reuse of any copyrighted component of this work in other works.}
\newcommand\copyrightnotice{%
\begin{tikzpicture}[remember picture,overlay]
 \node[anchor=south,yshift=10pt] at (current page.south) {\fbox{\parbox{\dimexpr\textwidth-\fboxsep-\fboxrule\relax}{\copyrighttext}}};
\end{tikzpicture}%
}

\title{\LARGE \bf
Linear Predictive Coding for Acute Stress Prediction from \\Computer Mouse Movements
}

\author{Lawrence H. Kim$^{1a}$, Rahul Goel$^{2a}$, Jia Liang$^{3a}$, Mert Pilanci$^{4}$, and Pablo E. Paredes$^{1}$
\thanks{$^{a}$These authors contributed equally}%
\thanks{$^{1}$Lawrence H. Kim and Pablo E. Paredes are with the Department of Psychiatry and Behavioral Sciences, Stanford University, Stanford, CA, USA
        {\tt\small \{lawkim,pparedes\}@stanford.edu}}%
\thanks{$^{2}$Rahul Goel was with the Department of Radiology, Stanford University, Stanford, CA, USA
        {\tt\small rahulg@alum.mit.edu}}%
\thanks{$^{3}$Jia Liang is with the Institute for Computational and Mathematical Engineering, Stanford University, Stanford, CA, USA
        {\tt\small jialiang@stanford.edu}}%
\thanks{$^{4}$Mert Pilanci is with the Department of Electrical Engineering, Stanford University, Stanford, CA, USA
        {\tt\small pilanci@stanford.edu}}}

\begin{document}



\maketitle
\thispagestyle{empty}
\pagestyle{empty}

\copyrightnotice

\begin{abstract}

Prior work demonstrated the potential of using the Linear Predictive Coding (LPC) filter to approximate muscle stiffness and damping from computer mouse movements to predict acute stress levels of users. Theoretically, muscle stiffness and damping in the arm can be estimated using a mass-spring-damper (MSD) biomechanical model. However, the damping frequency (i.e., stiffness) and damping ratio values derived using LPC were not yet compared with those from a theoretical MSD model. This work demonstrates that the damping frequency and damping ratio from LPC are significantly correlated with those from an MSD model, thus confirming the validity of using LPC to infer muscle stiffness and damping. We also compare the stress level binary classification performance using the values from LPC and MSD with each other and with neural network-based baselines. We found comparable performance across all conditions demonstrating LPC and MSD model-based stress prediction efficacy, especially for longer mouse trajectories.
\newline

\indent \textit{Clinical relevance}— This work demonstrates the validity of the LPC filter to approximate muscle stiffness and damping and predict acute stress from computer mouse movements.

\end{abstract}


\section{INTRODUCTION}

Stress is an instrumental factor for the emotional, cognitive, and physical well-being of people. A large corpus of research has demonstrated strong links between stress and a wide range of chronic health risks such as cardiovascular disease [1], diabetes [2], hypertension [2], obesity [2], and coronary artery disease [3]. Physiological reactions induced by stress are symptomatic of mental illnesses such as anxiety disorder and depression, which are a leading cause of suicides [4]. Chronic stress can also lead to mood swings, social isolation, and even reduction in academic achievements among adolescents [5].

Re-purposing data from everyday use human-machine interfaces have been proposed to continuously monitor and detect acute stress levels of individuals [6], [7]. While there are a plethora of data streaming into our smart devices that could be useful for stress prediction, movement data is of particular interest due to its privacy-preserving nature compared to other types of data such as app usage, messages, and location. Sun et al. demonstrated in the MouStress paper, the effectiveness of using computer mouse motion data to measure acute stress where the damping frequency (square of which is proportional to stiffness) values obtained from a linear predictive coding (LPC) filter, were larger under a stressed condition than a calm condition [6]. The steering wheel of a car simulator has also been successfully shown to detect acute stress, with data from only a few turns [7]. However, the parameters obtained from the LPC filter, in prior studies [6],[7], were not yet compared with corresponding parameters obtained from a theoretical mass-spring-damper (MSD) biomechanical model of the arms. This study aims to bridge that gap to help progress toward continuous stress monitoring using everyday mouse movement data.


\section{BIOMECHANICAL MODEL OF THE ARM}
\subsection{Mass-Spring-Damper (MSD) Model}
In the context of computer mouse interactions, what precedes clicking is a rapid, goal-directed skilled movement. This skilled movement is primarily open-loop in nature and is influenced by feedforward use of sensory information obtained in previous computer mouse movements and experience [9]–[13]. The dynamic behavior during any routine skilled movement is influenced by passive joint properties, initial motor commands, sensory feedback, mechanical constraints, and mechanics of the neuromuscular system, aka, biomechanics [14]. Research in biomechanics has shown that a rapid goal-directed movement can be modeled as a step response of a linear second-order system, i.e., a classic mass-spring-damper system [9]–[13]. Typically, smoothed raw movement data is fit to a simple open-loop feedforward model [12], [13].

\begin{figure}[]
\centering
    \framebox{\includegraphics[width=.97\linewidth]{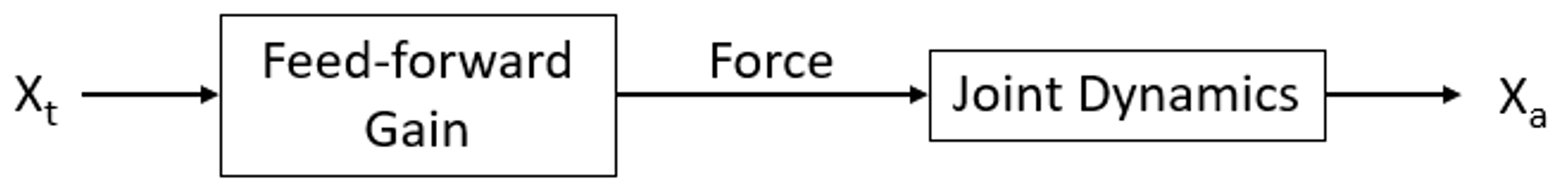}}
    \caption{Open-loop feedforward model for rapid goal-directed arm movement.}
    \label{fig:model}
\end{figure}

Fig. \ref{fig:model} shows the model describing the relationship between the target ($X_t$) and the actual ($X_a$) computer mouse position in the horizontal direction. The open-loop transfer function between $X_t$ and $X_a$ is given by Eq. 1 as:
$$
\frac{X_a(s)}{X_t(s)} = \frac{K_f}{Js^2+Bs+K} \eqno{(1)}
$$
where, $K_f$ represents the feedforward gain, J represents the moment of inertia of the arm, B is the viscous damping coefficient of the damper, and K is the stiffness of the spring. Eq. 1 can also be represented as: 
$$
\frac{X_a(s)}{X_t(s)} = \frac{K_p\omega^2}{s^2+2\omega\zeta s+\omega^2} \eqno{(2)}
$$
where, $K_p$ is the static gain, $\omega$ is the damping frequency, and $\zeta$ is the damping ratio. Parameters ($K_p$, $\omega$, $\zeta$) were estimated using an iterative time-domain identification technique called prediction-error minimization (“pem” from the System Identification Toolbox of MATLAB 2020b [Mathworks, Natick, MA]) that minimizes the cost function defined as the sum of squares of the difference between the actual and the simulated output. Normalized root mean square error was used to estimate the goodness of fit (GOF) of the model and is expressed as:
$$
100 \times \left(1-\frac{||X_a-X_s||}{||X_a-mean(X_a)||}\right) \eqno{(3)}
$$
where, $X_s$ is the simulated step response from the model.

\subsection{Linear Predictive Coding (LPC) Filtering Technique}
Linear predictive coding is a finite impulse response (FIR) filtering technique (Eq. 4) that builds a predictive model of future samples based only on linear combinations of observed signals from the past [15]. LPC model assumes an all-pole filter that can approximate systems where poles are dominating signal characteristics. It turns out that an ideal second-order system, such as the MSD system described earlier, which is an all-pole system in the Laplace domain, is similar in structure to that of a simple LPC model. That is, if we build an LPC model that best fits a series of samples, we can recover an approximation to the MSD parameters.

$$
H(z)=\frac{1}{A(z)},\ A(z)=\frac{E(z)}{X(z)}=1-\sum^p_{k=1}a_kz^{-k} \eqno{(4)}
$$
where H(z) is the system response, $a_k$ are the LPC coefficients, E(z) is the system approximation error, p is the order of the approximation. 

In our previous two studies [6] and [7], we used an LPC filter of order 4 that generates a sequence of 4 coefficients. Then, we constructed a 4th order polynomial from the 4 coefficients and considered the complex roots of it (that will exist in the case of an under-damped system). Then we estimated, the damping frequency ($\omega$) as the imaginary part of the complex root ($\omega = |I(r)|$) , and the damping ratio ($\zeta$) as the ratio of the complex root’s real part and its absolute value ($\zeta = \frac{|R(r)|}{||r||}$). We claimed that the equation of complex roots is similar to the characteristic polynomial of the 2nd order MSD system. Thus $\omega$ and $\zeta$ obtained from each of the two approaches (LPC and MSD) should be correlated. The primary aim of this study is to provide evidence towards that claim by showing the correlation between the parameters from the two approaches. It is important to note that the LPC approach is substantially simpler and lends itself to efficient real-time hardware and software implementations.


\section{METHODOLOGY}
\begin{figure}[]
\centering
    \framebox{\includegraphics[width=.8\linewidth]{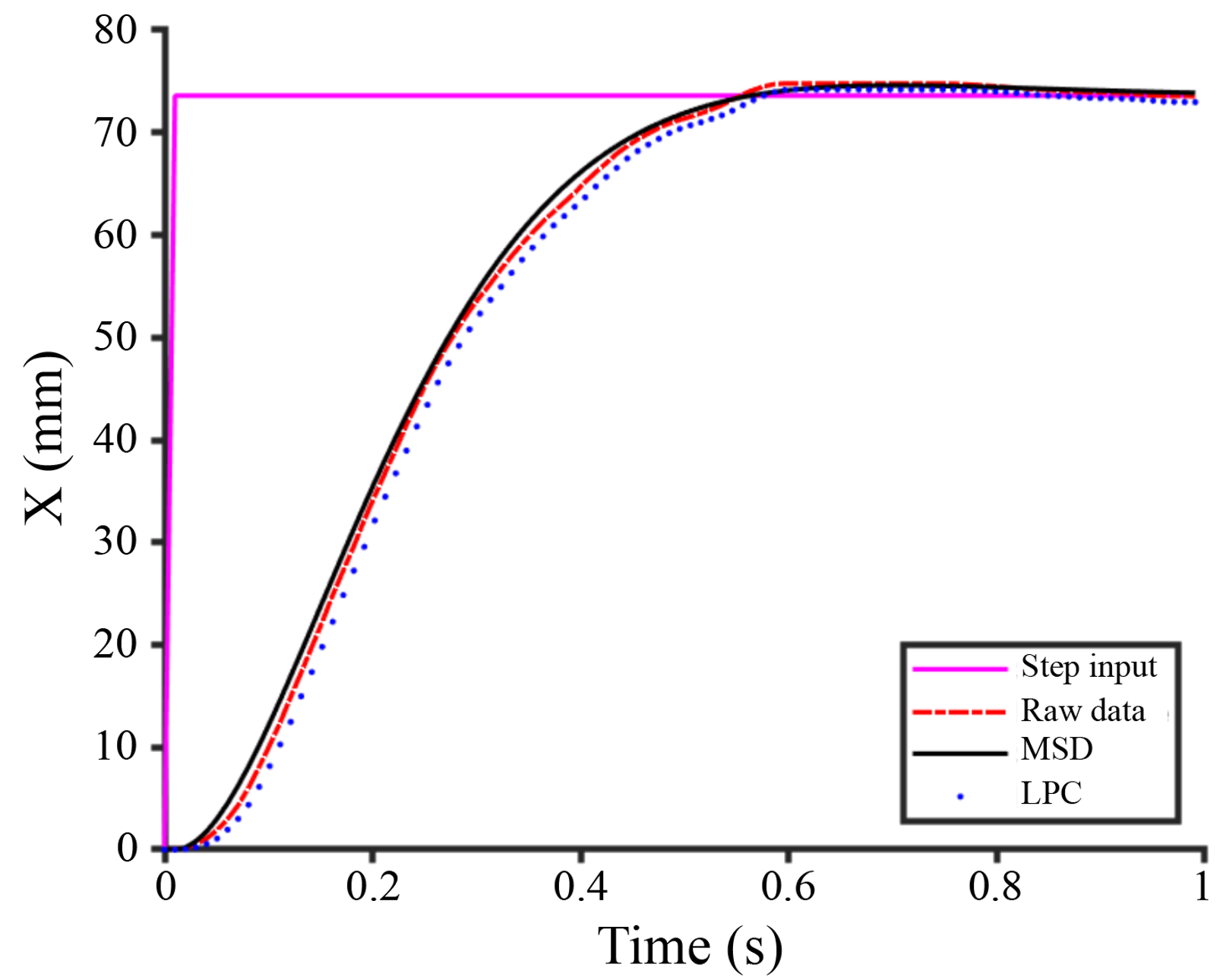}}
    \caption{An example of raw data compared with the responses from MSD and LPC.}
    \label{fig:rawdata}
\end{figure}

Using the dataset from the MouStress study [6], we computed the damping frequency, $\omega$, and damping ratio, $\zeta$, using the MSD model and LPC filter. Fig. \ref{fig:rawdata} shows an example of the resulting responses from MSD and LPC compared with the raw signal. Then, we computed the correlation between the $\omega$ and $\zeta$ from both methods after removing outliers and compared the classification performance when using LPC and MSD-based $\omega$ and $\zeta$, either as individual parameters or in combination.

\subsection{Dataset}
\begin{figure}[]
\centering
    \framebox{\includegraphics[width=.9\linewidth]{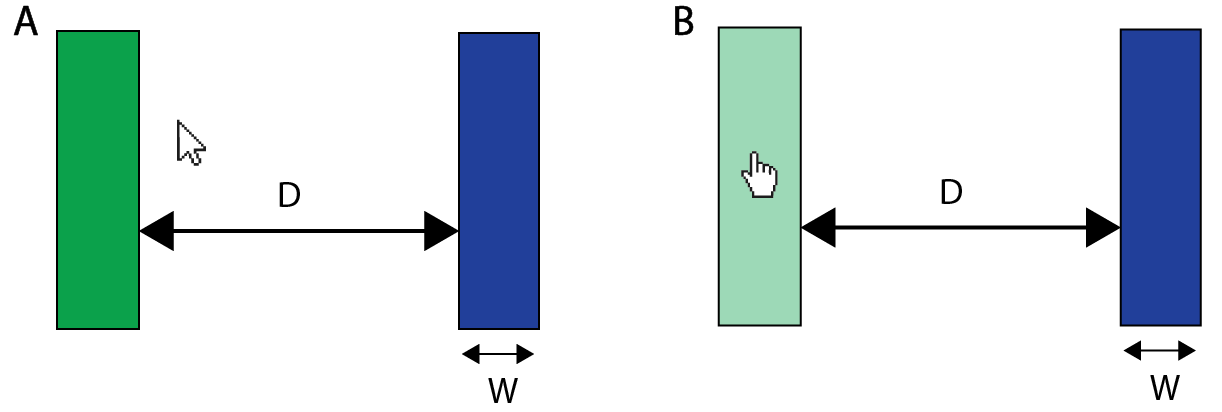}}
    \caption{Point-and-click task from MouStress [6]. a) Participants have to reach the green region and b) click it, which is then dimmed to provide feedback.}
    \label{fig:task}
\end{figure}

We used the data from the point-and-click task during the MouStress study [6], as clicking is the most frequent computer mouse event [8]. As the MSD model assumes rapid goal-directed movement, we primarily leverage the mouse movement data before the click to compute the $\omega$ and $\zeta$ using MSD and LPC. As described in the MouStress paper [6] and shown in Fig. \ref{fig:task}, the objective of the user was to move horizontally and click the two targets in succession (first, green on the left, and then, blue on the right) as quickly and accurately as possible. Since the primary movement direction was horizontal. we only looked at the x-axis movement data. Two task parameters, distance D (64px, 128px, 256px, 512px, 1024px) and width W (8px, 16px, 32px, 64px), were varied. Each of the N = 49 participants (26 female and 23 male, mean age = 20) performed 5 repetitions of the task with the same configuration under both stressed and calm conditions. The stressed conditions were elicited through 5 minutes of recursive mental math calculations. More details of the experimental design can be found in the MouStress paper [6].

\subsection{Correlation Analysis}
To find correlations between the $\omega$'s and $\zeta$'s obtained from each of the MSD and LPC approaches, we conducted Spearman’s rank correlations for different thresholds of GOF ranging from 0\% to 95\%.

\subsection{Binary Stress Classification}
We compared performance using eight different machine learning (ML) models. Six models were based on a standard ML technique, called the support vector machines (SVMs), and included models built with either individual parameters ($\omega$ or $\zeta$), or combination of parameters ($\omega$ and $\zeta$), derived from each of the MSD and LPC approaches. We also deployed two neural network-based ML methods, specifically Long Short-Term Memory (LSTM) and Convolutional Neural Network (CNN), using smoothed raw computer mouse movement data. Our goal was to compare the performance (in terms of accuracy) from MSD and LPC-based methods with neural network baselines, which do not leverage any domain knowledge or processing. To minimize the effects of inter-subject differences and the distance of point-and-click tasks on the performance of the tasks, we decided to build a different classifier per participant per distance of the point-and-click task. This led to each classifier having 40 data samples (20 trials for stressed condition and 20 trials for calm). Based on the data, we did not see significant influence of the width and thus did not take that into consideration. SVM  was chosen as some of our initial analyses comparing different standard ML classifiers like SVM, Decision Tree etc., with part of the data, consistently showed that SVM outperformed other standard ML classifiers.

For each trial, the dataset consisted of the x-position of the computer mouse obtained at a fixed frequency (2 kHz). As a result, input to the LSTM and CNN classifiers were time-series, and output was the binary stress condition (either stressed or calm). Since the same participant in each trial might have finished the point-and-click task at different times, the length of the time-series data was variable. Thus, we first tried to use a baseline LSTM network-based method. However, the direct application of the LSTM model on the raw data turned out not to be successful (based on non-decreasing validation loss). Further investigation of the raw data revealed that for all trials, the final value of x position of each time series was generally repeated  many times (which was obvious as the location of the target bar was fixed). As a result, the last several points in each time-series looked very similar regardless of the condition. Thus, we let the LSTM network learn from windowed time-series data with the end points removed. To implement this, we fixed a cutoff value (representing the last points of the new time-series) for all the point-and-click tasks for the same distances. The cutoff values were set to be 100 ms for 64x, 125 ms for 128x, 150 ms for 256x, 350 ms for 512x, and 500 ms for 1024x. To make a fair comparison, the same cutoff values applied to inputs of the LSTM model were also applied to that of the CNN model, which is described below.

\begin{figure}[]
\centering
    \framebox{\includegraphics[width=.5\linewidth]{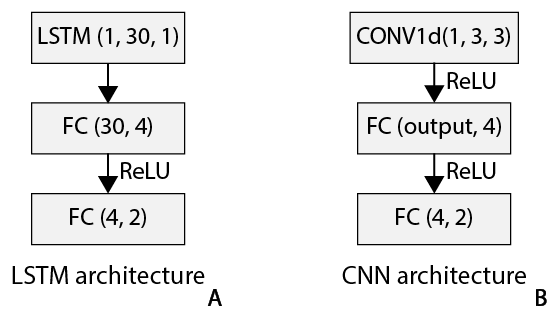}}
    \caption{Architecture of the methods based on A) LSTM and B) CNN. (FC = fully connected layer, ReLU = rectified linear unit)}
    \label{fig:architecture}
\end{figure}

Due to the scarcity of our data, the final version of the LSTM model we used consisted of only one LSTM layer followed by two fully-connected layers, and its architecture is shown in Fig. \ref{fig:architecture}A. The LSTM-based model was trained with Stochastic Gradient Descent Optimizer with cross-entropy loss as the loss function. The batch size was equal to 4, and the learning rate was fixed at 0.05. We split our data such that 80\% were used for training and validation, and the remaining 20\% for evaluation. The model was trained from scratch for 100 epochs, and the model with the best validation accuracy during training was saved and used to infer the test accuracy. This process was repeated 10 times for each subject and each distance.

The final architecture of the CNN model is shown in Fig. \ref{fig:architecture}B. The CNN model was trained exactly like the LSTM based model as described above with a learning rate of 0.001, but was trained for 500 epochs. Both the neural-network models were implemented in Pytorch 1.7.0.

For the standard ML models, the $\omega$ and $\zeta$ parameters obtained from each of MSD and LPC approaches were fit with the SVM classifier (either individually or in combination) with default hyperparameters and min-max normalization. SVM was imported from the sklearn library. The accuracy was obtained from averaging accuracies from 5-fold cross-validation for each iteration. The overall process (extraction of parameters with either MSD or LPC, fitting SVM, obtaining classification accuracy) described above was repeated 10 times, and the average results across the 10 iterations are reported. We also compare the classification performance from the LPC and MSD-based methods using paired t-tests after confirming normality via Kolmogorov-Smirnov test of normality.

\section{RESULTS \& DISCUSSION}
The means and standard errors of $\omega$ and $\zeta$ from MSD and LPC under calm and stressed conditions are computed after removing outliers, where $\zeta \leq 0$ or $\zeta > 100$, and are shown in Table \ref{tab:basic}. We see a significant ($p < 0.001$) increase in $\omega$ of the stressed condition relative to the calm condition. However, although the $\zeta$ from the LPC was significantly higher under stressed condition, than calm (as previously shown in [6]), the $\zeta$ from MSD tended to be lower in stressed condition than calm (as previously hypothesized in [6]).

\begin{table}[]
\caption{Mean (Standard Error) of $\omega$ and $\zeta$ from MSD and LPC models under calm and stressed conditions}
\label{tab:basic}
\begin{center}
\scriptsize 
    \begin{tabular}{|r|c|c|c|c|}
    \hline
    & \multicolumn{4}{c|}{$\omega$} \\
    \hline
    Model & calm & stressed & t(48) & p\\
    \hline
    MSD & 12.9 (0.5) & 14.4 (0.4) & 3.6 & $<0.001^*$\\
    LPC & 0.261 (0.002) & 0.268 (0.001) & 3.8 & $<0.001^*$\\
    \hline
    \hline
    & \multicolumn{4}{c|}{$\zeta$}\\
    \hline
    Model & calm & stressed & t(48) & p\\
    \hline
    MSD & 1.00 (0.03) & .97 (0.03) & 1.1 & 0.28\\
    LPC  & 0.5635 (0.0005) & 0.5652 (0.0005) & 3.2 & 0.002$^*$\\
    \hline
    \end{tabular}
\end{center}
\end{table}

\begin{figure}[]
\centering
    \framebox{\includegraphics[width=.9\linewidth]{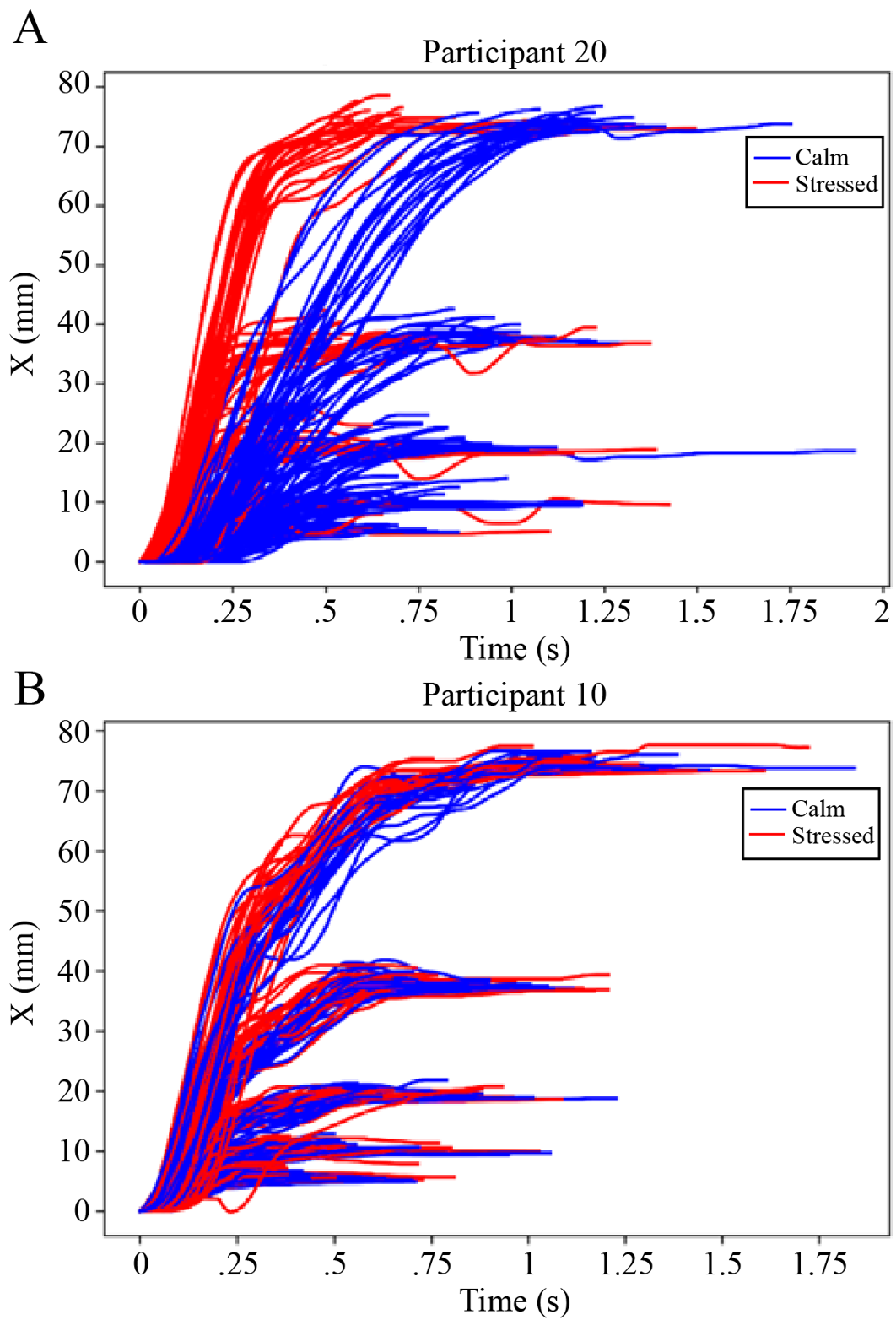}}
    \caption{Examples of mouse pointer trajectories from two participants. a) One with, and b) one without, visually clear boundaries between the stressed (Red) and calm (Blue) conditions.}
    \label{fig:example}
\end{figure}

Fig. \ref{fig:example} show examples of all the raw mouse motion trajectories from two representative participants under stressed and calm conditions. Fig. \ref{fig:example}A demonstrates computer mouse trajectories with clear visual distinction between the stressed and calm conditions. Such trends were noticed visually for 16 of the 49 participants when the distance between the bars was 1024x. Fig. \ref{fig:example}B shows a participant's data where no visually clear boundary could be drawn between the trajectories under stressed and calm conditions. 33 participants exhibited similar behavior.  

\subsection{Correlation}
\begin{figure}[]
\centering
    \framebox{\includegraphics[width=.97\linewidth]{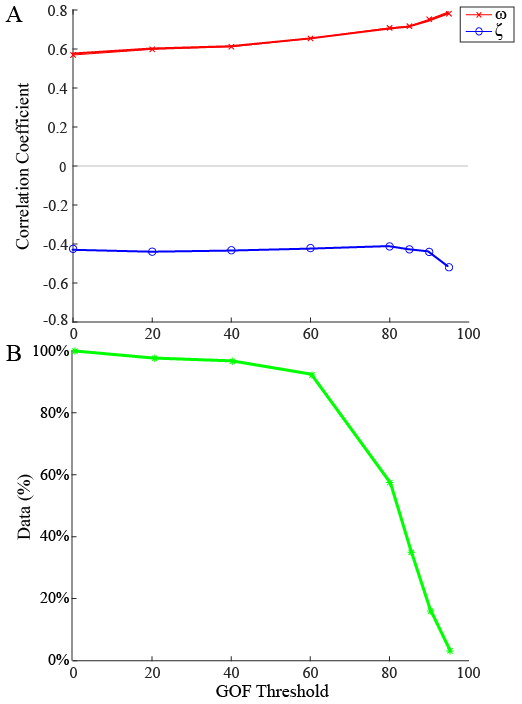}}
    \caption{A) The correlation coefficients between the $\omega$ and $\zeta$ from MSD and LPC for different thresholds of GOF are shown with red and blue lines, respectively. B) The green line indicates the percentage of the corresponding data according to different thresholds of GOF.}
    \label{fig:correlation}
\end{figure}

Fig. \ref{fig:correlation}A show the correlations between the damping frequencies and damping ratios from MSD and LPC for different thresholds of GOF. All correlations were statistically significant ($p<.001$). The correlation coefficient between the damping frequencies was strong and reached 0.7 with GOF of 80\% and went even higher for data with higher GOF. These results suggest that the damping frequency, as obtained from the LPC filter could be a very good proxy for the damping frequency, as obtained from a theoretical MSD model. For the damping ratios, we saw moderate correlations of approximately $-0.45$, even for data with GOF of 90\%. The negative correlation between the damping ratios suggests, that the damping ratio from LPC (although significantly different for stressed and calm conditions in our task), is not that good a proxy for damping ratio obtained in a traditional sense from a theoretical MSD model. Further, as the threshold for GOF reached 80\%, we saw a steeper decline in the percentage of data that met the requirement as shown in Fig. \ref{fig:correlation}B, where only around 55\% of the entire data met the 80\% GOF threshold criteria, typically common in system modeling studies.

\begin{table}[]
\caption{Comparison of binary stress classification's mean accuracy (standard errors), for the different ML models (chance was 50\%). Highest classification accuracy for each distance is highlighted in blue.}
\label{tab:accuracy}
\begin{center}
\resizebox{\linewidth}{!}{%
    \begin{tabular}{|r|c|c|c|c|c|c|c|c|}
    \hline
    \small
    \textbf{Distance} & CNN & LSTM & MSD & LPC & MSD & LPC & MSD & LPC\\
    & & & $\omega, \zeta$ & $\omega, \zeta$ & $\omega$ & $\omega$ & $\zeta$ & $\zeta$\\
    \hline
    64x	& 51.8\%	& 54.7\%	& 58.8\%	& 53.5\% & \textbf{\color{blue}{60.0\%}} & 54.6\% & 51.2\% & 57.1\%\\
    &(0.65\%) & (1.1\%)&(1.9\%) &(1.4\%) & (1.9\%)& (1.7\%)&(1.0\%) &(1.7\%)\\
    \hline
    128x&	53.6\%&	56.4\%&	56.4\%	&55.9\%& 57.1\% & 58.8\% & 52.3\% & \textbf{\color{blue}{58.9\%}}\\
    &(0.95\%) & (1.5\%)&(2.0\%) &(1.8\%) & (2.0\%)& (1.9\%)&(1.0\%) &(2.0\%)\\
    \hline
    256x	&55.1\%	&58.3\%	&59.1\%	&57.2\%& 59.5\% & \textbf{\color{blue}{62.3\%}} & 55.0\% & 60.4\%\\
    &(1.04\%) & (1.7\%)&(2.0\%) & (1.6\%)& (2.0\%)& (1.8\%)&(1.3\%) &(1.9\%)\\
    \hline
    512x	&60.1\%	&65.3\%	&60.0\%	&58.6\%& 60.7\% & \textbf{\color{blue}{67.2\%}} & 54.9\% & 64.4\%\\
    &(1.5\%) & (1.8\%)&(2.2\%) & (1.3\%)& (2.1\%)& (2.0\%)&(1.6\%) &(2.1\%)\\
    \hline
    1024x	&62.2\%	&67.3\%	&63.6\%	&61.4\%& 65.3\% & \textbf{\color{blue}{72.9\%}} & 56.0\% & 69.4\%\\
    &(1.6\%) & (2.0\%)&(2.3\%) & (1.4\%)& (2.2\%)& (1.8\%)&(1.6\%) &(1.9\%)\\
    \hline
    \textbf{Overall} & 56.6\%	&60.4\%	&59.6\%	&57.3\%& 60.5\% & \textbf{\color{blue}{63.2\%}} & 53.9\% & 62.0\%\\
    &(1.01\%) & (1.4\%)&(1.8\%) & (1.3\%)& (1.7\%)& (1.4\%) &(0.7\%) &(1.6\%)\\
    \hline
    \end{tabular}}
\end{center}
\end{table}

\subsection{Binary Acute Stress Classification}
Table \ref{tab:accuracy} shows the classification accuracies from the different ML models. The LPC $\omega$-based model had the highest overall accuracy, followed by LPC $\zeta$-based model, MSD $\omega$-based model, and LSTM-based model, respectively. Paired t-tests showed that LPC-based methods had significantly higher classification accuracies than MSD-based methods when either using $\omega$ ($p=.01$)  or $\zeta$ ($p<.001$). This result suggests that the LPC-based models can match or outperform MSD-based models and neural network-based baseline models for classifying acute binary stress levels. Specifically, the classifier based on the LPC-based $\omega$ yielded accuracy above 75\% for nearly half of the participants (23 of 49) for trials with 1024x distance. The relevance of LPC-based $\omega$ was also demonstrated in prior literature [6],[7] . In the future, it may be worth exploring whether using only the data samples above certain GOF (for MSD model) or below certain error variance (for LPC filter, and something not considered in this study) improves the accuracies of the MSD- and LPC- based classifiers, respectively. Further, it is possible that hyperparameter tuning in SVM would yield better classification accuracies. For the neural network-based classifiers, training with larger amount of data may yield higher classification accuracies.

It is also worth noting the reduction in accuracy observed with smaller distances between targets. A possible explanation for this reduction is that as the targets are closer together, there is less interaction from the larger muscles in the arm versus the hand. As we progress towards in-the-wild studies, it would be relevant to investigate optimal window lengths further to apply MSD or LPC-based estimators.

\section{CONCLUSION}
In this paper, we demonstrate that the parameters (especially damping frequency, aka stiffness) derived from an LPC filter are valid and a good proxy (aka strongly correlated) to those derived from a biomechanical MSD model of the human arm to predict binary acute stress levels of users based on their computer mouse movement data. We also compared stress classification performance using SVM to that of neural network-based models such as LSTM and CNN, and found that MSD and LPC-based models produced higher classification accuracies than LSTM and CNN-based models. This demonstrates the potential of LPC and MSD models in predicting the acute stress levels of users from their computer mouse movement data to enable continuous stress monitoring. In the future, we plan to explore using a combination of neural network-based approaches with MSD or LPC, applying those methods to analyze in-the-wild data, and combining physiological signals to further improve the classification performance.

\addtolength{\textheight}{-12cm}   









\end{document}